\documentclass[prl,aps,amsfonts,twocolumn,showpacs]{revtex4}

\newcommand{\be}[1]{\begin{equation} #1 \end{equation}}

\newcommand{\ba}[2]{\left(\begin{array}{#1}#2\end{array}\right)}

\newtheorem{theorem}{Theorem}

\newcommand{\C}{{\mathbb C}}
\def\qed{\leavevmode\unskip\penalty9999 \hbox{}\nobreak\hfill
     \quad\hbox{\leavevmode  \hbox to.77778em{%
               \hfil\vrule   \vbox to.675em%
               {\hrule width.6em\vfil\hrule}\vrule\hfil}}
     \par\vskip3pt}

\begin{document}

\title{Four qubits can be entangled in nine different ways.}
\author{F.Verstraete$^{1,2}$,
J.Dehaene$^2$, B.De Moor$^2$ and H.Verschelde$^1$}
%\author{Frank Verstraete\cite{FV} \and Jeroen Dehaene \and Bart De Moor}
\affiliation{$^1$Ghent University, Department of Mathematical
Physics and Astronomy
Krijgslaan 281 (S9), B-9000 Gent, Belgium\\
$^2$Katholieke Universiteit Leuven, Department of Electrical
Engineering, Research Group SISTA Kasteelpark Arenberg 10, B-3001
Leuven, Belgium}

\begin{abstract}
We consider a single copy of a pure four-partite state of qubits
and investigate its behaviour under the action of stochastic local
quantum operations assisted by classical communication (SLOCC).
This leads to a complete classification of all different classes
of pure states of four-qubits. It is shown that there exist nine
families of states corresponding to nine different ways of
entangling four qubits. The states in the generic family give rise
to GHZ-like entanglement. The other ones contain essentially 2- or
3-qubit entanglement distributed among the four parties. The
concept of concurrence and 3-tangle is generalized to the case of
mixed states of 4 qubits, giving rise to a seven parameter family
of entanglement monotones. Finally, the SLOCC operations
maximizing all these entanglement monotones are derived, yielding
the optimal single copy distillation protocol.
\end{abstract}
\pacs{03.65.Ud}

\maketitle

One of the open questions in the field of quantum information
theory is to understand the different ways in which multipartite
systems can be entangled. As the concept of entanglement is
related to the non-local properties of a state, local quantum
operations cannot affect the intrinsic nature of entanglement. It
is therefore natural to define equivalence classes of states
generated by the group of reversible SLOCC
operations\cite{BPRST01,DVC00}. In this letter we are concerned
with SLOCC operations on one copy of a state, which means that we
are considering actions under LOCC operations on one copy of a
state without imposing that they can be achieved with unit
certainty. Two states belonging to the same class are able to
perform the same QIT-tasks, although with a different probability.

In the case of a single copy of an entangled pure state of two
qubits, it is well known that it can be converted to the singlet
state by SLOCC operations \cite{LP97}. In the case of three
entangled qubits, it was shown \cite{DVC00,AJDV00,VDD01d} that
each state can be converted by SLOCC operations either to the
GHZ-state $(|000\rangle +|111\rangle)/\sqrt{2}$, or to the W-state
$(|001\rangle +|010\rangle +|100\rangle)/\sqrt{3}$, leading to two
inequivalent ways of entangling three qubits. The GHZ-state is
generally considered as the state with the genuine 3-partite
entanglement, while the W-state has the peculiar property of
having the maximal expected amount of two-partite entanglement if
one party is traced out \cite{DVC00}. In this letter we extend
these results to the case of four qubits. Furthermore the widely
celebrated entanglement measures concurrence \cite{Wo98} and
3-tangle \cite{CKW00}, characterizing the amount of genuine two-
and three-qubit entanglement, are generalized to the case of four
qubits, giving rise to a 7 parameter family of entanglement
monotones. The SLOCC filtering operations maximizing all these
entanglement monotones are derived, and it is shown that these are
the unique operations \cite{VDD01c} (up to local unitaries)
bringing a state into a locally stochastic form (i.e. bringing all
local density operators equal to the identity). Following Gisin
\cite{Gi98}, we claim that these operations maximize the true
4-partite entanglement.

Interestingly, we found that there exist eight families of pure
4-qubit states that cannot be brought into local stochastic form
by finite SLOCC operations. These states do have the peculiar
property that they have the maximal amount of 2- and/or 3-qubit
entanglement shared between all 4 parties. In some sense their
entanglement is maximally robust against the loss of one or two
qubits.

An interesting feature about entanglement that emerges out of the
results of this letter is the fact that a quantum state has only a
finite susceptibility for entanglement. This will be illustrated
by the fact that the operations maximizing the true 4-partite
entanglement are precisely the operations that destroy all local
correlations (i.e. the local density operators are made
stochastic) and that also destroy the 3-partite entanglement (i.e.
the 3-tangle of the states obtained by tracing out one party
becomes equal to zero). The states having maximal 2- or 3-partite
entanglement shared among the four parties on the other hand are
exactly the states having zero genuine 4-partite entanglement
(i.e. the 4-concurrences are all equal to zero).

Before developing the mathematical formalism, it should be noted
that the study of states of four qubits is particularly
interesting as the current experimental state of the art allows to
entangle four photons \cite{PDGWZ01,WZ01,LHB01} or ions
\cite{Sa00}. Furthermore SLOCC operations can relatively easily be
implemented on photons, and it is therefore of interest to
implement the optimal SLOCC operations such as to yield a state
with maximal 4-partite entanglement.

This letter is organized as follows. First we derive a simple way
of determining whether two pure 4-qubit states are connected by
local unitary operations. Next some advanced linear algebra is
used to determine the orbits generated by SLOCC operations. This
leads to nine different families of states, corresponding to nine
essentially different ways of entangling four qubits, although
only one family is generic. This analysis gives rise to seven
independent entanglement monotones characterizing the 4-partite
entanglement. Finally the optimal SLOCC operations are derived
such as to maximize all these entanglement monotones.

Let us now first consider the problem to determine whether two
pure 4-qubit states are equivalent up to local unitary operations.
Therefore the following accident in Lie-group theory can be
exploited:
\[SU(2)\otimes SU(2)\simeq SO(4)\]
Here $SO(4)$ denotes the family of real orthogonal matrices with
determinant equal to $1$. More specifically, it holds that
$\forall U_1,U_2\in SU(2): T(U_1\otimes U_2)T^\dagger\in SO(4)$
where {\small
\be{T=\frac{1}{\sqrt{2}}\ba{cccc}{1&.&.&1\\.&i&i&.\\.&-1&1&.\\i&.&.&-i}\label{T}.}}
A pure state of four qubits is parameterized by a four index
tensor $\psi_{i_i i_2 i_3 i_4}$ with $i_j\in\{1,2\}$. This tensor
can be rewritten as a $4\times 4$ matrix $\tilde{\psi}$ by
concatenating the indices $(i_1,i_2)$ and $(i_3,i_4)$. Next we
define the matrix R as \be{R=T\tilde{\psi}T^\dagger.\label{R}} It
is then straightforward to show that a local unitary
transformation $|\psi'\rangle =U_1\otimes U_2\otimes U_3\otimes
U_4|\psi\rangle$ results in a transformation $R'=O_1 R O_2$ with
$O_1,O_2 \in SO(4)$ and $O_1=T(U_1\otimes U_2) T^\dagger$,
$O_2=T(U_3\otimes U_4)^T T^\dagger$. A normal form under local
unitary operations can now be imposed as follows: make the $(1,1)$
entry of $R$ real by multiplying the whole matrix with the
appropriate phase, and use $O_1$ and $O_2$ to diagonalize the real
part of $R$ through the unique real singular value decomposition.
This procedure eliminates all 13 degrees of freedom of the local
unitary operations, and two states are therefore equivalent up to
local unitary operations iff they have the same normal form.

Next we move to the central problem of this letter, namely
characterizing the local orbits generated by SLOCC operations of
the form \be{|\psi'\rangle=A_1\otimes A_2\otimes A_3\otimes
A_4|\psi\rangle\label{SLOCC}} with $\{A_i\}$ full rank and
therefore invertible $2\times 2$ matrices. There is no restriction
in choosing $\{A_i\}\in SL(2,\C)$, and then a new useful accident
arises: \be{SL(2,\C)\otimes SL(2,\C)\simeq SO(4,\C).} $SO(4,\C)$
denotes the non-compact group of complex orthogonal matrices $O^T
O=I_4$. Again it holds that $\forall A,B \in SL(2,\C):
T(A_1\otimes A_2)T^\dagger \in SO(4,\C)$ with $T$ given in
equation (\ref{T}), and SLOCC operations therefore correspond to
left and right multiplication of $R$ (\ref{R}) with complex
orthogonal matrices. The challenge is now to exploit the two times
12 degrees of freedom of these complex orthogonal matrices to
bring $A$ into an unique normal form with maximal 8 real degrees
of freedom left. This will be possible using some advanced
techniques of linear algebra.

We will now state a technical theorem that is a generalization of
the singular value decomposition to complex orthogonal matrices.
\begin{theorem}\label{OSVD}
Given a complex $n\times n$ matrix $R$, then there always exist
complex square orthogonal matrices $O_1$ and $O_2$ such that
$R'=O_1RO_2$ is a unique direct sum of blocks of the form:\\
1. $m\times m$ blocks of the form $\left(\lambda_jI_m+S_m\right)$
being symmetric Jordan blocks (see for example \cite{HJ85} 4.4.9),
and $\lambda_j$ is a complex parameter (note that the
case $m=1$ corresponds to the scalar case).\\
2. $m\times m$ blocks consisting of an upper left $(m_1+1)\times
m_1$ part being the matrix obtained by taking the odd rows and
even columns of an $(2m_1+1)\times (2m_1+1)$ symmetric Jordan
block, and a lower right $(m-m_1-1)\times (m-m_1)$ part being the
transpose of the matrix obtained by taking the odd rows and even
columns of a $(2(m-m_1)-1)\times (2(m-m_1)-1)$ symmetric Jordan
block.
\end{theorem}
Proof: Consider the $2n\times 2n$ complex symmetric matrix
\begin{equation}P=\ba{cc}{0&R\\R^T&0}\label{PQ}.\end{equation} Due to theorem 5 in ch.XI of
\cite{Ga59}, there exists a complex orthogonal $Q$ such that
$P=QP'Q^T$ with $P'$ a direct sum of symmetric $m\times m$ Jordan
blocks $J_i$ with eigenvalue $\lambda_i$. Next we observe that
whenever $[v_1;v_2]$ ($v_1$ and $v_2$ both have $n$ rows such that
$[v_1;v_2]$ has $2n$ rows) is the eigenspace of $P$ corresponding
to a symmetric Jordan block $J_i$, then $[v_1;-v_2]$ is the
eigenspace of $P$ corresponding to a Jordan block $-J_i$. Due to
the uniqueness of the Jordan canonical decomposition, these
eigenspaces will be either linearly independent (this holds for
example for sure if the corresponding eigenvalue is different from
zero), or equal to each other (which implies that the
corresponding eigenvalue is equal to zero). If the first case
applies, both $v_1$ and $v_2$ are orthogonal matrices.

The second degenerated case however is more difficult. In this
case, it holds that $[v_1;v_2]=[v_1;-v_2]Q$ for some orthogonal
$Q$. Let us first calculate the standard non-symmetric Jordan
canonical form $\tilde{J}$ of the symmetric Jordan block with
eigenvalue 0: $J=U^\dagger\tilde JU$ with $U$ unitary and
symmetric. If we define $[x_1;x_2]=[v_1;v_2]U^\dagger$ and
$\tilde{Q}=U^\dagger Q U$, the following identities hold:
$\tilde{Q}^T{\rm Sip}\tilde{Q}={\rm Sip}$,
$\tilde{Q}\tilde{J}=-\tilde{J}\tilde{Q}$ and
$[x_1;x_2]^T[x_1;x_2]={\rm Sip}$ (the matrix ${\rm Sip}$ is
defined as the matrix permuting all vectors $[x_1,x_2\cdots x_n]$
to $[x_n,x_{n-1}\cdots x_1]$). The conditions on $\tilde{Q}$ imply
that
$\tilde{Q}$ is equal to the matrix $\tilde{Q}_{ij}=\pm(-1)^i\delta_{ij}$.% Dit komt omdat Q=diag(1,-1,1,-1,...)*R met R=[a b c ...\\0 a b c ...\\0 0 a b c ...] en SipQ^TSipQ\simeq R^2=I en dat kan alleen als b=c=...=0!
Therefore $[x_1;x_2]$ is either of the form
\begin{equation}\ba{c}{x_1\\x_2}=\ba{cccccc}{a_1&0&b_1&0&c_1&\cdots\\0&a_2&0&b_2&0&\cdots}\label{x1x2}\end{equation}
or
\begin{equation}\ba{c}{x_1\\x_2}=\ba{cccccc}{0&a_1&0&b_1&0&\cdots\\a_2&0&b_2&0&c_2&\cdots}\label{x2x1}.\end{equation}
Due to the constraint $[x_1;x_2]^T[x_1;x_2]={\rm Sip}$, the row
dimension of $[x_1;x_2]$ and therefore of $J$ has to be odd, as
otherwise the upper rightmost entry cannot be equal to 1.
Retransforming to the original picture with the unitary $U$, it
holds that this structure is preserved, and the eigenspace
$[v_1;v_2]$ is of a form (\ref{x1x2}) or (\ref{x2x1}).

As the dimension of a $J_i$ giving rise to the degenerated case
has to be odd, it is compulsory that there is an even number of
degenerated cases (indeed, the non-degenerate cases give rise to
two times a similar block and the total dimension is even). More
precisely, for each $[v_1;v_2]_j$ of the form (\ref{x1x2}), there
has to exist a $[v_1;v_2]_k$ of the form (\ref{x2x1}) (eventually
of different dimension). The eigenstructure of such pairs of
degenerate cases can then be brought into the form
\[\ba{cccccccccccc}{a_1^i& b_1^i\cdots
&a_1^k&b_1^k&\cdots&0&0&\cdots&0&0&\cdots\\\cdots&0&0&\cdots&0&0&\cdots
a_2^i& b_2^i\cdots &a_2^k&b_2^k&\cdots}\] by right multiplication
with a permutation matrix $W$. The effect on $J_i$ and $J_k$ is to
transform them as
\[W^T\ba{cc}{J_i&0\\0&J_k}W=\ba{cccc}{0&0&K_i&0\\0&0&0&K_k^T\\K_i^T&0&0&0\\0&K_k&0&0}\]
where $K_\nu$ represents the matrix obtained by taking the odd
rows and even columns of the symmetric Jordan block $J_\nu$.

Collecting all the pieces, it is now easily verified that the
canonical form obtained is exactly of the form stated in the
theorem. This completes the proof.\qed

Due to the equivalence of $SL(2,\C)\otimes SL(2,\C)$ and
$SO(4,\C)$, the normal forms arising in  the above lemma will
immediately yield a natural representative state for each class of
4-qubit states connected by SLOCC operations. The normal form
encodes the genuine non-local properties of the state, while the
SLOCC operators needed to bring the state into normal form
characterize the local information. The following classification
is obtained:
\begin{theorem}
A pure state of 4 qubits can, up to permutations of the qubits, be
transformed into one of the following 9 families of states by
determinant 1 SLOCC operations (\ref{SLOCC}): {\small
\begin{eqnarray*}G_{abcd}&=&\frac{a+d}{2}(|0000\rangle
+|1111\rangle)+\frac{a-d}{2}(|0011\rangle
+|1100\rangle)\\
&&\hspace{.1cm}+\frac{b+c}{2}(|0101\rangle
+|1010\rangle)+\frac{b-c}{2}(|0110\rangle +|1001\rangle)\\
L_{abc_2}&=&\frac{a+b}{2}(|0000\rangle
+|1111\rangle)+\frac{a-b}{2}(|0011\rangle +|1100\rangle)\\
&&\hspace{.5cm}+c(|0101\rangle +|1010\rangle)+|0110\rangle \\
L_{a_2b_2}&=&a(|0000\rangle +|1111\rangle)+b(|0101\rangle+|1010\rangle)\\
&&\hspace{.5cm}+|0110\rangle +|0011\rangle \\
L_{ab_3}&=&a(|0000\rangle
+|1111\rangle)+\frac{a+b}{2}(|0101\rangle
+|1010\rangle)\\
&&\hspace{.5cm}+\frac{a-b}{2}(|0110\rangle +|1001\rangle)\\
&&\hspace{.5cm}+\frac{i}{\sqrt{2}}(|0001\rangle +|0010\rangle
+|0111\rangle
+|1011\rangle)\\
L_{a_4}&=&a(|0000\rangle +|0101\rangle +|1010\rangle
+|1111\rangle)\\
&&\hspace{.5cm}+(i|0001\rangle +|0110\rangle -i|1011\rangle)\\
L_{a_20_{3\oplus\bar{1}}}&=&a(|0000\rangle
+|1111\rangle)+(|0011\rangle +|0101\rangle +|0110\rangle)\\
L_{0_{5\oplus\bar{3}}}&=&|0000\rangle +|0101\rangle +|1000\rangle
+|1110\rangle \\
L_{0_{7\oplus\bar{1}}}&=&|0000\rangle +|1011\rangle +|1101\rangle
+|1110\rangle\\
L_{0_{3\oplus\bar{1}}0_{3\oplus\bar{1}}}&=&|0000\rangle+|0111\rangle\end{eqnarray*}}
The complex parameters $a,b,c,d$ are the unique eigenvalues of P
(\ref{PQ}) with non-negative real part, and the indices
$L_{\alpha\beta\cdots}$ are representative for the Jordan block
structure of P (e.g. $L_{a_20_{3\oplus\bar{1}}}$ means that the
eigenstructure of $P$ consists of two $2\times 2$ Jordan blocks
with eigenvalues $a$ and $-a$, and a degenerated pair of dimension
respectively 3 and 1).
\end{theorem}
{\em Proof:} If theorem \ref{OSVD} is applied to a $4\times 4$
$R$, it is easily checked that 12 different families arise where a
family is defined as having Jordan and degenerated Jordan blocks
of specific dimension. Note however that the orthogonal matrices
obtained by application of the theorem can have determinant equal
to $-1$, while the SLOCC operations correspond to an orthogonal
matrix with determinant $+1$; this is however not a problem as
these operations correspond to SLOCC operations followed by a
permutation of the qubits ($1\leftrightarrow 2$) or
($3\leftrightarrow 4$). One can proceed by checking that
permutations of qubits ($2\leftrightarrow 3$) or
($1\leftrightarrow 4$) transform different families into each
other. It is indeed true that $R=J_1(a)\oplus J_1(b)\oplus
K_{3\oplus \bar{1}}$ transforms into $R'=J_2(a)\oplus J_2(b)$ if
qubit 2 and 3 are permuted. This also happens in the case
$J_1(a)\oplus K_{5\oplus\bar{1}}\hspace{.2cm}\rightarrow J_4(a)$.
Moreover it can be shown that $J_1(a)\oplus K_{3\oplus\bar{3}}$ is
equivalent to $J_1(a)\oplus J_3(0)$. Therefore only 9 essentially
different normal forms are retained.\qed

A generic pure state of 4 qubits can always be transformed to the
$G_{abcd}$ state. This state is peculiar in the sense that all
local density operators, obtained by tracing out all parties but
one, are proportional to the identity. As shown in \cite{VDD01c},
this is the unique state (up to local unitary operations) with
this property of all states connected by SLOCC operations. In the
light of the results of Gisin \cite{Gi98} and Nielsen about
majorization \cite{Ni99,NK01}, we claim that this is the state
with maximal 4-partite entanglement on the complete orbit
generated by SLOCC operations: the more entanglement, the more
local entropy. In a later section this argument will be made hard
by showing that a whole class of entanglement monotones are indeed
maximized for the locally stochastic state.

It is interesting to note that the 3-tangle \cite{CKW00} of the
mixed states obtained by tracing out one party of this $G_{abcd}$
state is always equal to zero. Indeed, if the right-unitary matrix
U {\small
\begin{eqnarray*} U&=&\frac{1}{\sqrt{2(1+|\beta |^2)}}\ba{cccc}{1 & \beta & 1
&-\beta \\ \beta & 1 &-\beta & 1 }\\
\beta &=&\sqrt{-q+\sqrt{q^2-r}}\\
q&=&8\,{a}^{2}{d}^{2}+8\,{b}^{2}{c}^{2}-4\,{a}^{2}{b}^{2}-4\,{a}^{2}{c}^{2}-4\,{d}^{2}{b}^{2}-4\,{d}^{2}{c}^{2}\\
r&=&(a^2-d^2)(b^2-c^2)\end{eqnarray*}} is applied to the $8\times
2$ matrix {\small
\[\ba{cccccccc}{a+d&.&.&a-d&.&b+c&b-c&.\\.&b-c&b+c&.&a-d&.&.&a+d}^T\]}
being the square root of the density operator obtained by tracing
out the first qubit, 4 3-qubit W-states are obtained. If we define
the mixed 3-tangle \cite{VDD01c} as the convex roof of the square
root of the 3-tangle, this quantity is clearly equal to zero.
Therefore the SLOCC operations maximizing the 4-partite
entanglement result in a loss of all true 3-partite entanglement.
This is reminiscent to the case of 3 qubits where the 2-qubit
state obtained by tracing out one particle of a GHZ-state is
separable.

Let us next discuss some specific examples. A completely separable
state belongs to the family $L_{abc_2}$ with $a=b=c=0$. If only
two qubits are entangled, an EPR state arises belonging to the
family $L_{a_2b_2}$ with $a=b=0$. A state consisting of two
EPR-pairs belongs to $G_{abcd}$ with $(a=1;b=c=d=0)$ or $a=b=c=d$,
depending on the permutation. The class
$L_{0_{3\oplus\bar{1}}0_{3\oplus\bar{1}}}$ consists of all 3-qubit
GHZ states accompanied with a separable qubit, while the 3-qubit
W-state belongs to the family $L_{a_20_{3\oplus\bar{1}}}$ with
$a=0$.

The 4-qubit $|\Phi_4$-state \cite{BR01} belongs to the generic
family, while the 4-qubit W-state $(|0001\rangle +|0010\rangle
+|0100\rangle +|1000\rangle)/2$ belongs to the family $L_{ab_3}$
with $a=b=0$. This W-state can be shown to have a mixed 3-tangle
equal to zero, but has a concurrence of 1/2 when whatever two
qubits are traced out. On the contrary the state
$L_{O_{7\oplus\bar{1}}}$ has all concurrences equal to zero if two
qubits are traced out. This state is completely symmetric in the
permutation of the qubits 2,3 and 4. It has the property of having
a mixed 3-tangle equal to 1/2 if particle 2,3 or 4 is traced out.
This can be proven by considering the $8\times 2$ "square root"
{\small
\[\frac{1}{2}\ba{cccccccc}{1&.&.&.&.&.&.&1\\.&.&.&1&.&.&1&.}^T.\]} Some
straightforward calculations show that the average square root of
the 3-tangle of the vectors obtained by multiplying this matrix
with whatever $2\times n$ right-unitary matrix is equal to 1/2.
Similar arguments show that only three-qubit W-type entanglement
($\tau=0$) is retained if the first qubit is traced out.

The state $L_{0_{5\oplus\bar{3}}}$ is somehow a hybrid of both the
4-qubit W-state and  $L_{O_{7\oplus\bar{1}}}$. Again a mixed
3-tangle of 1/2 is obtained if qubit 2,3 or 4 is traced out, a
mixed 3-tangle equal to zero if qubit 1 is traced out, but now the
mixed state obtained by tracing out qubit 1 and (3 or 4) has a
concurrence equal to 1/2, while the other concurrences vanish.

Another interesting state belongs to the family $L_{a_4}$ with
$a=0$: $|\psi\rangle=(|0001\rangle +|0110\rangle
+|1000\rangle)/\sqrt{3}$. Its mixed 3-tangle equals 2/3 in the
case of tracing out qubit 1 or 4 and vanishes otherwise. Moreover
the concurrence vanishes everywhere if 2 qubits are traced out
except in the case of tracing out qubit 2 and 3, resulting in a
concurrence of 2/3.

After this zoological survey, let us next move on to the topic of
entanglement monotones. The complex eigenvalues of P (\ref{PQ}),
given by $\pm(a,b,c,d)$, are the only invariants under all
determinant 1 SLOCC operations (note that an eigenvalue 0 is
associated to the degenerated Jordan blocks). In \cite{VDD01c} it
was proven that all real positive functions of the parameters of a
pure state that are linearly homogeneous in $\rho$ and remain
invariant under determinant 1 SLOCC operations, are entanglement
monotones (in the case of mixed states they are defined by the
convex roof formalism). Therefore all real positive homogeneous
functions of $(a^2,b^2,c^2,d^2)$ are entanglement monotones, such
as \[
M_\alpha(\psi)=|a^\alpha+b^\alpha+c^\alpha+d^\alpha|^{2/\alpha}.
\]
Taking into account one degree of freedom due to the phase, this
gives rise to a seven-parameter family of entanglement monotones.
All these entanglement monotones are maximized by the operations
making the density matrix locally stochastic \cite{VDD01c}
(meaning that the identity is obtained when all qubits but one are
traced out). The optimal single-copy distillation procedure for a
generic pure state is therefore to implement the SLOCC operations
bringing it into its normal form $G_{abcd}$. This is in complete
accordance with the results of Nielsen on majorization
\cite{Ni99}. Note that all the other normal forms can only be
brought into the local stochastic normal form by a filtering
procedure whose probability of success tends to zero
\cite{VDD01c}.

In summary, we have identified all different families of pure
states of 4 qubits generated by SLOCC operations. Only one family
is generic, and all states in it can be made locally stochastic by
SLOCC operations. The same SLOCC operations represent the optimal
single-copy distillation protocol. The eight other families
correspond to states having some kind of degenerated 4-partite
entanglement and are the 4-partite generalizations of the
3-partite W-state.

\bibliographystyle{unsrt}
%\bibliography{c:/frank/articles/frank}

\end{document}